\documentstyle[12pt,preprint]{aastex}
\newcommand{\bjdtdb}{${\rm {BJD_{TDB}}}$}
\newcommand{\feh}{{\left[{\rm Fe}/{\rm H}\right]}}
\newcommand{\teff}{{T_{\rm eff}}}

\newcommand{\msun}{${\rm M}_\Sun$}
\newcommand{\rsun}{${\rm R}_\Sun$}

\newcommand{\mj}{${\,{\rm M}_{\rm J}}$}
\newcommand{\rj}{${\,{\rm R}_{\rm J}}$}
\newcommand{\fave}{\langle F \rangle}
\newcommand{\fluxcgs}{10$^9$ erg s$^{-1}$ cm$^{-2}$}

\begin{document}

\title{KELT-2Ab: A Hot Jupiter Transiting the Bright (V=8.77) Primary Star of a Binary System}

\author{Thomas G.\ Beatty\altaffilmark{1},
  Joshua Pepper\altaffilmark{2},
  Robert J.\ Siverd\altaffilmark{2},
  Jason D.\ Eastman\altaffilmark{3,4},
  Allyson Bieryla\altaffilmark{5},
  David W. Latham\altaffilmark{5},
  Lars A.\ Buchhave\altaffilmark{6,7},
  Eric L.\ N.\ Jensen\altaffilmark{8},
  Mark Manner\altaffilmark{9},
  Keivan G. Stassun\altaffilmark{2,10},
  B.\ Scott Gaudi\altaffilmark{1},
  Perry Berlind\altaffilmark{5},
  Michael L.\ Calkins\altaffilmark{5},
  Karen Collins\altaffilmark{11},
  Darren L. DePoy\altaffilmark{12},
  Gilbert A.\ Esquerdo\altaffilmark{5},
  Benjamin J.Fulton\altaffilmark{3},
  G\'{a}bor F\H{u}r\'{e}sz\altaffilmark{5},
  John C.\ Geary\altaffilmark{5},
  Andrew Gould\altaffilmark{1},
  Leslie Hebb\altaffilmark{2},
  John F. Kielkopf\altaffilmark{11},
  Jennifer L. Marshall\altaffilmark{12},
  Richard Pogge\altaffilmark{1},
  K.Z.Stanek\altaffilmark{1},
  Robert P.\ Stefanik\altaffilmark{5},
  Rachel Street\altaffilmark{3},
  Andrew H.\ Szentgyorgyi\altaffilmark{5},
  Mark Trueblood\altaffilmark{13},
  Patricia Trueblood\altaffilmark{13},
  \&
  Amelia M.\ Stutz\altaffilmark{14,15}
}

\altaffiltext{1}{Department of Astronomy, The Ohio State University, 140 W.\ 18th Ave., Columbus, OH 43210}
\altaffiltext{2}{Department of Physics and Astronomy, Vanderbilt University, Nashville, TN 37235}
\altaffiltext{3}{Las Cumbres Observatory Global Telescope Network, 6740 Cortona Drive, Suite 102, Santa Barbara, CA 93117}
\altaffiltext{4}{Department of Physics Broida Hall, University of California, Santa Barbara, CA 93106}
\altaffiltext{5}{Harvard-Smithsonian Center for Astrophysics, 60 Garden Street, Cambridge, MA 02138}
\altaffiltext{6}{Niels Bohr Institute, University of Copenhagen,Juliane Maries vej 30, 21S00 Copenhagen, Denmark}
\altaffiltext{7}{Centre for Star and Planet Formation, Geological Museum, {\O}ster Voldgade 5, 1350 Copenhagen, Denmark}
\altaffiltext{8}{Department of Physics and Astronomy, Swarthmore College, Swarthmore, PA 19081}
\altaffiltext{9}{Spot Observatory, Nunnelly, TN 37137}
\altaffiltext{10}{Department of Physics, Fisk University, Nashville, TN 37208}
\altaffiltext{11}{Department of Physics \& Astronomy, University of Louisville, Louisville, KY 40292}
\altaffiltext{12}{Department of Physics \& Astronomy, Texas A\&M University, College Station, TX 77843}
\altaffiltext{13}{Winer Observatory, Sonoita, AZ 85637}
\altaffiltext{14}{Max Planck Institute for Astronomy, Heidelberg, Germany}
\altaffiltext{15}{Department of Astronomy \& Steward Observatory, University of Arizona, Tucson, AZ 85721} 

\shorttitle{KELT-2Ab}
\shortauthors{Beatty et al.}

\begin{abstract}
We report the discovery of KELT-2Ab, a hot Jupiter transiting the bright (V=8.77) primary star of the HD 42176 binary system.  The host is a slightly evolved late F-star likely in the very short-lived ``blue-hook'' stage of evolution, with $\teff=6148\pm48{\rm K}$, $\log{g}=4.030_{-0.026}^{+0.015}$ and $\feh=0.034\pm0.78$. The inferred stellar mass is  $M_*=1.314_{-0.060}^{+0.063}$\msun\ and the star has a relatively large radius of $R_*=1.836_{-0.046}^{+0.066}$\rsun. The planet is a typical hot Jupiter with period $4.11379\pm0.00001$ days and a mass of $M_P=1.524\pm0.088$\mj\ and radius of $R_P=1.290_{-0.050}^{+0.064}$\rj. This is mildly inflated as compared to models of irradiated giant planets at the $\sim$4 Gyr age of the system.  KELT-2A is the third brightest star with a transiting planet identified by ground-based transit surveys, and the ninth brightest star overall with a transiting planet. KELT-2Ab's mass and radius are unique among the subset of planets with $V<9$ host stars, and therefore increases the diversity of bright benchmark systems. We also measure the relative motion of KELT-2A and -2B over a baseline of 38 years, robustly demonstrating for the first time that the stars are bound. This allows us to infer that KELT-2B is an early K-dwarf. We hypothesize that through the eccentric Kozai mechanism KELT-2B may have emplaced KELT-2Ab in its current orbit. This scenario is potentially testable with Rossiter-McLaughlin measurements, which should have an amplitude of $\sim$44 m s$^{-1}$. 
\end{abstract}


\section{Introduction}
Individual giant planets transiting bright main sequence stars remain of prime scientific interest. While the multitude of hot Jupiters orbiting fainter (i.e., $V>9$) stars provides an opportunity to learn about the statistical properties of giant planets, it is the hot Jupiters around the bright stars which provide us with specific information about planetary interiors and atmospheres \citep[see, e.g.,][]{winn2010a}. Indeed, since their discovery, all of the transiting hot Jupiters orbiting bright ($V<9$) stars have been observed repeatedly from space and the ground for precisely this reason \citep{seager2010}. Since there are currently only five transiting giant planets in this magnitude range, discovering even one more substantially increases the opportunities for these important, detailed, follow-up observations.  

The Kilodegree Extremely Little Telescope (KELT) North transit survey is designed to find precisely these planets. KELT-North uses a small aperture telescope with a very wide field of view to observe a strip in declination that covers approximately 40\% of the Northern sky. The combination of a small light-collecting area and a wide field of view for KELT-North was chosen to efficiently survey all of the dwarfs stars between $8<V<10$ in our footprint. We specifically chose this magnitude range to cover the gap that exists between radial velocity surveys ($V\lesssim8.5$) and other transit surveys ($V\gtrsim10$). 

The KELT-North survey has been in operation since 2006, and we have been actively generating and vetting planet candidates since 2011 April. In this letter we describe the discovery and characterization of a hot Jupiter transiting the bright primary component of the HD 42176 binary system, which we hereafter refer to as the KELT-2 system.  

\section{Discovery and Follow-up Observations}
The KELT-North telescope consists of an Apogee AP16E (4K$\times$4K 9$\mu$m pixels) CCD camera attached to a Mamiya 645 camera lens with 42mm aperture and 80mm focal length (f/1.9). The resultant field of view is $26^\circ\times26^\circ$ at roughly 23\arcsec per pixel. The telescope uses a Kodak Wratten \#8 red-pass filter and the resultant bandpass resembles a widened Johnson-Cousins R-band. The telescope is located at Winer Observatory in Sonoita, AZ. \cite{pepper2007} give a more detailed description of the telescope and instrumentation.

KELT-2 is in KELT-North survey field 04, which we monitored from 2006 October 27 to 2011 March 31, collecting a total of 7,837 observations on 136,702 stars in the field. We reduced the raw survey data using a custom implementation of the ISIS image subtraction package \citep{alard1998,alard2000}, combined with point-spread fitting photometry using {\sc daophot} \citep{stetson1987}. Using the Tycho-2 proper motions and a reduced proper motion cut based on \cite{collier2007}, we identified 29,345 putative dwarf and sub-giant stars within field 04 for further post-processing and analysis. We applied the trend filtering algorithm \citep[TFA,][]{kovacs2005} to each dwarf star lightcurve to remove systematic noise, followed by a search for transit signals using the box-fitting least squares algorithm \citep[BLS,][]{kovacs2002}. For both TFA and BLS we used the versions found in the {\sc vartools} package \citep{hartman2008}. A more detailed description of our data reduction, post-processing, and candidate selection can be found in \cite{siverd2012}.        
 
KELT-2 passed our cuts based on the results of the BLS transit search, and was promoted to by-eye candidate vetting. The KELT-North lightcurve showed a strong transit-like signal from a relatively isolated star, with no significant power in either an Analysis of Variance \citep{schwarzenberg1989,devor2005} or a Lomb-Scargle periodogram \citep{lomb1976,scargle1982,press1989,press1992}. To check for blending, we also examined if the center of light moved off the position of the star during transit, which it did not. At this point matching to the CCDM and WDS catalogs revealed that KELT-2 was a suspected close visual binary system. We considered KELT-2 a strong planet candidate, and passed it on for follow-up observations. All of the KELT-2 data can be downloaded from the KELT-North website.\footnote{www.astronomy.ohio-state.edu/keltnorth/data/kelt2/}

\subsection{Follow-up Spectroscopy}
Our follow-up spectroscopy was collected using the Tillinghast Reflector Echelle Spectrograph (TRES), on the 1.5m Tillinghast Reflector at the Fred L. Whipple Observatory (FLWO) at Mt. Hopkins, AZ. We specifically targeted KELT-2A for our follow-up; the typical seeing for TRES is 1\arcsec.5, which allowed us to exclude most of the light from KELT-2B 2\arcsec.3 away. The spectra had a resolving power of R=44,000, and were extracted following the procedures described by \cite{buchhave2010}.

Between UT 2012 February 1 and UT 2012 May 4 we obtained 18 spectra of KELT-2A. We have excluded one observation taken on the night of UT 2012-02-07 which suffered from higher than normal contamination from the Moon. The top panel of Figure 1 shows our radial velocity observations phased and overplotted to our final orbital fit. The residuals to this fit and the bisector spans are shown in the middle and bottom panels, respectively. Note that the bisectors do not show any phase structure, and the 24.42 m s$^{-1}$ RMS in the bisector values is 6.5 times smaller than the velocity amplitude of our final phased orbit: 161.1 m s$^{-1}$. This indicates that we are likely seeing true reflex Doppler motion in the spectra of the parent star, KELT-2A. The residuals to our orbital fit have a 22.74 m s$^{-1}$ RMS.

\subsection{Follow-up Photometry}
We obtained follow-up transit photometry of KELT-2A between late February and late 2012 March. Figure 2 shows the phased transit lightcurves. For all of the transit observations the observers deliberately defocused their telescopes to minimize the noise arising from interpixel variations coupled with pointing drift, and to prevent saturation. This meant that in none of our follow-up transit photometry were KELT-2A and -2B resolved. We find the same transit depths in our $g$, $i$ and $z$ follow-up lightcurves, which helps rule out stellar false positives.\footnote{After accounting for the small ($\sim$1\%) but wavelength-dependent effect of KELT-2B's added light on the transit depths.}

A complete transit in \emph{g} on UT 2012 February 20 and a partial transit with ingress in \emph{z} on UT 2012 March 29 were observed from Hereford Arizona Observatory (HAO), a private facility in southern Arizona. We observed with a 0.35m Meade Schimdt-Cassegrain equipped with a Santa Barbara Instrument Group (SBIG) ST-10XME camera using a Kodak KAF-3200E 2K$\times$1.5K CCD. The field of view of the observations was 26\arcmin.9$\times$18\arcmin.1 with a pixel scale of 0\arcsec.74 per pixel. We also used an SBIG AO-7 tip-tilt image stabilizer to hold the target steady on the detector.

The Peter van de Kamp Observatory at Swarthmore College observed a partial transit with egress on UT 2012 March 21 in \emph{i}. The Observatory uses a 0.6m RCOS telescope with an Apogee U16M 4K$\times$4K CCD, giving a 26\arcmin$\times$26\arcmin\ field of view. Using 2$\times$2 binning, this gives a pixel scale of 0\arcsec.76 per pixel. Fog arrived immediately after egress during these observations, so the out-of-transit baseline is shorter than usual.

We observed a partial transit with egress on UT 2012 March 25 with KeplerCam on the 1.2m telescope at FLWO. KeplerCam has a single 4K$\times$4K Fairchild CCD with a pixel scale of 0\arcsec.366 per pixel, for a total field of view of 23\arcmin.1$\times$23\arcmin.1. Observations were obtained in \emph{z}. Thin clouds passed overhead for most of the egress; we have removed these observations from the lightcurve. The data were reduced using a light curve reduction pipeline outlined in \cite{carter2011}, which uses standard IDL routines.

In addition to transit lightcurves, Bruce Gary obtained absolute photometry of the KELT-2 system in \emph{B}, \emph{V}, \emph{r}, \emph{i} and \emph{z} filters from HAO on UT 2012 April 16.

\section{Planetary and Stellar System Fitting}
KELT-2 is a binary system that is listed in several catalogs\footnote{e.g., 2MASS, NOMAD, UCAC3, GSC and ASCC} as a single star with a combined brightness of $V=8.71$. As part of the fitting process described below we have determined that the KELT-2 system is composed of a $V=8.77\pm0.01$, F7V primary in orbit with a $V=11.9\pm0.2$, K2V secondary. The two stellar components are presently separated by 2\arcsec.3. While the two stars have been assumed to be associated based on their proximity \citep{couteau1975}, imaging gathered as part of our follow-up observations and discussed in Section 3.1 proves they are bound. Table 1 lists general system and catalog information for the KELT-2 system.

The presence of KELT-2B as an unresolved component in all of our follow-up photometry -- both in the transit lightcurves and the broadband magnitudes -- complicated our fitting. We wished to remove the light from B to deblend the transit lightcurves, and to measure the actual apparent $V$ magnitude of KELT-2A. The latter is important as it allowed us to make an independent estimate of KELT-2A's radius in conjunction with the Hipparcos parallax.

The available magnitudes of the two stars \citep{couteau1975}, provided insufficient information on the properties of KELT-2B for us to accurately remove its light. We therefore performed an iterative round of Markov Chain Monte Carlo (MCMC) fitting to the lightcurves and radial velocities (RVs) to determine the properties of KELT-2A, followed by fitting the spectral energy distribution (SED) of the binary to identify the contribution from KELT-2B. We conducted two rounds of these fits: a first MCMC fit to the lightcurves with no deblending and at the Hipparcos-derived distance to provide initial parameters, followed by an SED fit to determine the properties of KELT-2B. The SED fitting also provided us with an estimate of the reddening to KELT-2. We then used the results of this first SED fit to deblend the transit lightcurves and reran the MCMC fitting. At this point we also included the Hipparcos parallax as a constraint and added the system distance as an additional free parameter. The refined system parameters from the second MCMC fit then fed into a second SED fitting. This SED fit gave parameters for KELT-2B within 1\% of the first SED fit, and so we judged the fitting process to have converged.

To perform the MCMC fits to the transit lightcurves and the radial velocity data we used the {\sc exofast} package \citep{eastman2012}. This is a suite of routines that performs MCMC fits to lightcurves and RV data simultaneously. {\sc exofast} also fits the properties of the parent star using the relations from \cite{torres2010}. We modified {\sc exofast} to include the distance prior from Hipparcos and account for KELT-2B's flux. The inclusion of the Hipparcos-derived distance allowed us to also fit for the distance to KELT-2. For each transit, we fit a unique airmass detrending coefficient and baseline flux and subtracted the band-dependent companion flux. We assumed a constant period in our final fit.

As initial inputs for {\sc exofast} we used values for the effective temperature ($6146\pm50\ {\rm K}$), surface gravity ($4.03\pm0.1$), projected rotational velocity ($9.0\pm2.0\ {\rm km\ s^{-1}}$) and metallicity ($0.06\pm0.08$) of KELT-2A derived from our follow-up spectroscopy. These were determined using the Spectral Parameter Classification (SPC) procedure \citep{buchhave2012}. SPC cross-correlates synthetic spectra created from a grid of Kurucz model atmospheres against the observed TRES spectra to estimate the spectral parameters and their uncertainties.
 
For our SED fitting, we fit combined model spectra of KELT-2A and B to our measured \emph{B}, \emph{V}, \emph{g}, \emph{i} and \emph{z} magnitudes for the KELT-2 system. We assigned KELT-2A the physical parameters from the {\sc exofast} MCMC fit and considered several possible spectra for KELT-2B over a range of effective temperatures from 4200K to 5300K, all with the same metallicity as KELT-2A and a fixed $\log g=4.65$. We calculated a radius for each temperature and $\log g$ using the \cite{torres2010} relations. We also included reddening as a free parameter. We constructed our model spectra with the NextGen model atmospheres \citep{hauschildt1999}.

Table 2 shows the stellar and planetary parameters derived from this iterative procedure. We include the results for a fit with eccentricity fixed to be zero, and another with eccentricity included as another free parameter. While formally the eccentricity for this second fit is non-zero ($e=0.182_{-0.084}^{+0.081}$), note that the values for $e\cos \omega_*$ and $e\sin \omega_* $ are nearly exactly zero. We view this eccentricity as insignificant, considering the Lucy-Sweeney bias \citep{lucy1971}. We therefore have adopted the $e\equiv0$ results as the true system parameters. Additionally, in both fits there is a small but non-zero slope in the RVs ($0.63\pm0.24$ m s$^{-1}$ day$^{-1}$). If real, this is too high to be accounted for by the binary orbit. An object in an edge-on circular orbit around KELT-2A should show a maximum RV slope of 0.63 m s$^{-1}$ day$^{-1}$ ($m_p/2$\mj) ($a/$AU)$^{-2}$. 

The planet KELT-2Ab has a mass of $M_P=1.524\pm0.088$\mj\ and radius of $R_P=1.290_{-0.050}^{+0.064}$\rj. The radius is mildly inflated as compared to \cite{baraffe2008}'s model for an similarly irradiated solar-composition giant planet (1 to 2\mj) at 3 to 4 Gyr (see \S 4), which would predict a radius of $\sim$1.1\rj. Nevertheless, KELT-2Ab is in the middle of the observed radii for planets of this mass, and follows the trend of radius inflation versus planetary effective temperature described by \cite{laughlin2011}.

The star KELT-2A is a slightly evolved F7 dwarf. We find $\teff=6148\pm48{\rm K}$, $\log{g}=4.030_{-0.026}^{+0.015}$ and $\feh=0.034\pm0.78$, with an inferred mass $M_*=1.314_{-0.060}^{+0.063}$\msun\ and a relatively large radius $R_*=1.836_{-0.046}^{+0.066}$\rsun.  The distance we determine to the KELT-2 system, $128.6_{-4.2}^{+5.2}$ pc, is 1.2$\sigma$ farther away than the $110\pm15$ pc determined using the Hipparcos parallax alone. 

For KELT-2B the results of our SED fits place the star as a K2V, with $\teff=4850\pm150{\rm K}$. Using the temperature-mass and temperature-radius relations from \cite{torres2010} -- and assuming $\log g=4.65$ -- we estimate KELT-2B has a mass of $M_*\approx0.78$\msun\ and radius of $R_*\approx0.70$\rsun. 

\subsection{The KELT-2 Binary System}
As part of our follow-up observations we obtained five high resolution \emph{z} images of the KELT-2 system from Spot Observatory, Nunnelly, TN on UT 2012 April 10 that resolved the two components. Spot Observatory is equipped with an RC Optics 0.6m telescope and an SBIG STL-6303E camera. For these observations, we used an SBIG AO-L transmissive adaptive optics system, which enabled a FWHM of $\sim$1\arcsec.35.

In each of the five images we were able to measure the relative positions and fluxes of KELT-2A and -2B. We did so by fitting a two dimensional Gaussian to KELT-2A, subtracting it, and then doing the same for KELT-2B. Figure 3 shows our measurements for the position of KELT-2B with respect to KELT-2A, along with those from \cite{couteau1975} and \cite{argue1992}. We also plot the position KELT-2B would have if it started at the 1973.96 average position and were unassociated with KELT-2A. For this we assumed KELT-2B had the average proper motion of K-giants between $11<V<13$ and within $1^\circ$ of the galactic coordinates of KELT-2A. We determined this average proper motion using the Besancon galaxy model \citep{robin2003}, which gave $\mu_\alpha=0.04\pm0.27$ mas yr$^{-1}$ and $\mu_\delta=-0.26\pm0.39$ mas yr$^{-1}$. Note that under this assumption the effect of parallax on the relative position of the two stars at these epochs is 10 times smaller than our uncertainties. We calculated the uncertainties on our average position and the 1973.96 average position by taking the standard deviation of the individual positions. From our observations, we exclude the possibility that these two stars are unassociated at the 8-$\sigma$ level.  

Our measurement of the relative positions of KELT-2A and B places KELT-2B at $\Delta\alpha=1\arcsec.19\pm0.03$ and $\Delta\delta=1\arcsec.95\pm0.05$ relative to KELT-2A for 2012.27. This corresponds to a position angle of $328^\circ.6$ and a separation of 2\arcsec.29. From the distance we determine this is a projected separation of $295\pm10$ AU. At this semimajor axis, the binary orbital period would be $\sim$3,500 years.

\section{Discussion}
The final planetary parameters for KELT-2Ab place it in a region of mass-radius parameter space that is already well populated by other hot Jupiters. What is noteworthy about this system is the brightness of the primary star, the primary's evolutionary state and the presence of a K2V common proper motion companion.

At an apparent magnitude of V=8.77, KELT-2A is the ninth brightest star with a known transiting planet, and the third brightest discovered by a ground-based transit survey\footnote{According to the Extrasolar Planets Encyclopedia at the date of writing.}. This makes KELT-2Ab an excellent candidate for both space- and ground-based follow-up work. In terms of the bright ($V<9$) transiting planets, KELT-2Ab, at 1.52\mj, allows access to a region of the mass-radius diagram otherwise unprobed by the known bright systems. HD 209458b, HD 149026b and HD 189733b are all less than 1.15\mj, while HD 17156b and HAT-P-2b are both over 3\mj. WASP-33b does not have a well-constrained mass, and 55 Cnc e and Kepler-21b are both super-Earths.  

The star KELT-2A itself is in an interesting region of parameter space. Comparison to the Yonsei-Yale isochrones \citep{demarque2004} with our determined temperature, surface gravity, and metallicity suggests that KELT-2A has just left the main sequence and its convective core is in the process of halting hydrogen fusion and the star is transitioning to shell burning. This so-called `blue-hook' transition \citep[see, e.g.,][]{exter2010}, which occurs immediately prior to the star's rapid transition across the Hertzsprung gap to the base of the red giant branch, only lasts a few tens of millions of years. Assuming that the isochrones and our inferred stellar properties are correct in placing KELT-2A on the `blue-hook,' we thus find a remarkably precise system age of $3.968\pm0.010$ Gyr. 

The existence of the stellar companion KELT-2B raises the intriguing possibility that KELT-2Ab migrated inward to its present location through the eccentric Kozai mechanism \citep{lithwick2011}. If this were true, then the orbit of the planet is likely misaligned with the spin axis of KELT-2A. Interestingly, the effective temperature of KELT-2A (6151K) places this system near the proposed dividing line between cool aligned and hot misaligned planetary systems noted by \cite{winn2010}. Future Rossiter-McLaughlin measurements of the system's spin-orbit alignment should provide insight into the efficiency of any mechanisms that might align planets around the cooler stars. We would expect, from Equation (6) of \cite{gaudi2007}, the amplitude of the Rossiter-McLaughlin anomaly to be $\sim$44 m s$^{-1}$. 

Given the brightness and spectral type of KELT-2A, it is interesting to ask why this system was not observed by any of the RV surveys for exoplanets. In addition to KELT-2A being fainter than most of the targets for RV surveys, it may be that the RV surveys did not examine the KELT-2 system because it is listed as a binary in many of the available catalogs.

\acknowledgments
We particularly thank Bruce Gary for obtaining the HAO observations. Early work on KELT-North was supported by NASA Grant NNG04GO70G. Work by B.S.G., J.D.E., and T.G.B.\ was partially supported by NSF CAREER Grant AST-1056524. E.L.N.J.\ gratefully acknowledges the support of the National Science Foundation's PREST program, which helped to establish the Peter van de Kamp Observatory through grant AST-0721386. J.A.P.\ and K.G.S.\ acknowledge support from the Vanderbilt Office of the Provost through the Vanderbilt Initiative in Data-intensive Astrophysics. K.G.S.\ and L.H.\ acknowledge the support of the National Science Foundation through PAARE grant AST-0849736 and AAG grant AST-1009810. K.A.C. is supported by a Kentucky Space Grant Consortium Graduate Fellowship. The TRES and KeplerCam observations were obtained with partial support from the Kepler Mission through NASA Cooperative Agreement NNX11AB99A with the Smithsonian Astrophysical Observatory (PI: D.W.L.).

This work has made use of NASA's Astrophysics Data System, the Exoplanet Orbit Database at exoplanets.org, the Extrasolar Planet Encyclopedia at exoplanet.eu \citep{schneider2011}, and the Washington Double Star Catalog maintained at the U.S. Naval Observatory.

\begin{deluxetable}{llclc}
\tablecaption{KELT-2 System Properties.}
\tablewidth{0pt}
\tabletypesize{\small}
\startdata
\hline
\hline
\colhead{~~~Parameter} & \colhead{Units} & \colhead{Value} & \colhead{Source} & \colhead{Ref.}\\
\hline
Names & & HD 42176 & & \\
 & & HIP 29301 & & \\
 & & WDS J06107+3057 & & \\
 & & BD+30 1138 & & \\
 & & CCDM J06107+3057 & & \\
 & & TYC 2420-899-1 & & \\
$\alpha_{\mathrm{J2000}}$ & & 06:10:39.35 & Hipparcos & 1\\
$\delta_{\mathrm{J2000}}$ & & +30:57:25.7 & Hipparcos & 1\\
$B$ (system)  & & $9.25\pm0.03$ & This Letter & \\
$V$ (system)  & & $8.70\pm0.01$ & This Letter & \\
$r$ (system)  & & $8.57\pm0.01$ & This Letter & \\
$i$ (system)  & & $8.48\pm0.01$ & This Letter & \\
$z$ (system)  & & $8.49\pm0.02$ & This Letter & \\
$J$ (system)  & & $7.669\pm0.021$ & 2MASS & 2\\
$H$ (system)  & & $7.417\pm0.026$ & 2MASS & 2\\
$K$ (system)  & & $7.346\pm0.031$ & 2MASS & 2\\
$\mu_\alpha$ & Proper motion in R.A. (mas yr$^{-1}$)\dotfill & $21.2\pm1.4$ & NOMAD & 3\\
$\mu_\delta$ & Proper motion in decl. (mas yr$^{-1}$)\dotfill & $-1.5\pm0.8$ & NOMAD & 3\\
$A_V$ & Visual Extinction\dotfill & $0.06\pm0.02$ & This Letter & \\
\enddata
\tablecomments{1=\cite{vanleeuwen2007}, 2=\cite{cutri2003}, 3=\cite{zacharias2005}}
\end{deluxetable}

\begin{deluxetable}{lccc}
\tablecaption{Median values and 68\% confidence intervals for the Physical and Orbital Parameters of the KELT-2A System.}
\tablewidth{0pt}
\tabletypesize{\small}
\startdata
\hline
\hline
\colhead{~~~Parameter} & \colhead{Units} & \colhead{Value ($e\neq0$)} & \colhead{Value ($e\equiv0$, adopted)}\\
\hline
\sidehead{RV Parameters:}
                                ~~~$M_{*}$\dotfill &Mass (\msun)\dotfill & $1.317_{-0.059}^{+0.062}$ & $1.314_{-0.060}^{+0.063}$\\
                              ~~~$R_{*}$\dotfill &Radius (\rsun)\dotfill & $1.842_{-0.073}^{+0.090}$ & $1.836_{-0.046}^{+0.066}$\\
                   ~~~$\log{g}$\dotfill &Surface gravity (cgs)\dotfill   & $4.027_{-0.035}^{+0.028}$ & $4.030_{-0.026}^{+0.015}$\\
                   ~~~$\teff$\dotfill &Effective temperature (K)\dotfill & $6147\pm50$               & $6148\pm48$\\
~~~$v\sin(i)$\dotfill &Projected rotation velocity (km s$^{-1}$)\dotfill & $9.0\pm2.0$               & $9.0\pm2.0$\\
                                  ~~~$\feh$\dotfill &Metallicity\dotfill & $0.038\pm0.077$           & $0.034\pm0.78$\\
                                   ~~~$d$\dotfill &distance (pc)\dotfill & $128.9_{-5.8}^{+6.5}$     & $128.6_{-4.2}^{+5.2}$\\
\sidehead{Planetary Parameters:}
                                   ~~~$e$\dotfill &Eccentricity\dotfill & $0.182_{-0.084}^{+0.081}$  & $\equiv0$\\
        ~~~$\omega_*$\dotfill &Argument of periastron (degrees)\dotfill & $-180_{-100}^{+110}$       & $\equiv90$\\
                                  ~~~$P$\dotfill &Period (days)\dotfill & $4.1137917\pm0.00001$      & $4.1137913\pm0.00001$\\
                           ~~~$a$\dotfill &Semi-major axis (AU)\dotfill & $0.05508\pm0.00085$        & $0.05504\pm0.00086$\\
                                 ~~~$M_{P}$\dotfill &Mass (\mj)\dotfill & $1.492\pm0.096$            & $1.524\pm0.088$\\
                               ~~~$R_{P}$\dotfill &Radius (\rj)\dotfill & $1.307_{-0.064}^{+0.077}$  & $1.290_{-0.050}^{+0.064}$\\
      ~~~$T_{\mathrm{eq}}$\dotfill &Equilibrium temperature (K)\dotfill & $1714_{-31}^{+37}$         & $1712_{-20}^{+28}$\\
                   ~~~$\fave$\dotfill &Incident flux (\fluxcgs)\dotfill & $1.90\pm0.15$              & $1.951_{-0.089}^{+0.13}$\\
\sidehead{RV Parameters:}
       ~~~$T_C$\dotfill &Time of inferior conjunction (\bjdtdb)\dotfill & $2455974.60361_{-0.00085}^{+0.00081}$ & $2455974.60338_{-0.00083}^{+0.00080}$\\
               ~~~$T_{P}$\dotfill &Time of periastron (\bjdtdb)\dotfill & $2455975.0\pm1.1$                     & $\equiv T_C$\\
                 ~~~$K$\dotfill &RV semi-amplitude (m s$^{-1}$)\dotfill & $160.5\pm8.7$                         & $161.1_{-8.0}^{+7.6}$\\
                    ~~~$M_P\sin{i}$\dotfill &Minimum mass (\mj)\dotfill & $1.491\pm0.096$                       & $1.523\pm0.088$\\
                           ~~~$\gamma_{0}$\dotfill &km s$^{-1}$\dotfill & $47.5\pm0.2$                          & $47.5\pm0.2$\\
    ~~~$\dot{\gamma}$\dotfill &RV slope (m s$^{-1}$ day$^{-1}$)\dotfill & $0.63\pm0.26$                         & $0.63\pm0.24$\\
                                ~~~$e\cos(\omega_*)$\dotfill & \dotfill & $-0.04\pm0.15$                        & $\equiv0$\\
                                ~~~$e\sin(\omega_*)$\dotfill & \dotfill & $0.00\pm0.15$                         & $\equiv0$\\                   
\sidehead{Primary Transit Parameters:}                                                          
~~~$R_{P}/R_{*}$\dotfill &Radius of the planet in stellar radii\dotfill & $0.0730\pm0.0018$            & $0.0723\pm0.0018$\\
           ~~~$a/R_*$\dotfill &Semi-major axis in stellar radii\dotfill & $6.43_{-0.26}^{+0.21}$       & $6.464_{-0.21}^{+0.098}$\\
                          ~~~$i$\dotfill &Inclination (degrees)\dotfill & $88.5_{-1.3}^{+1.0}$         & $88.56_{-1.3}^{+0.98}$\\
                               ~~~$b$\dotfill &Impact parameter\dotfill & $0.16_{-0.11}^{+0.15}$       & $0.16_{-0.11}^{+0.14}$\\
                             ~~~$\delta$\dotfill &Transit depth\dotfill & $0.00533\pm0.00027$          & $0.00522\pm0.00026$\\
              ~~~$\tau$\dotfill &Ingress/egress duration (days)\dotfill & $0.0151_{-0.0019}^{+0.0024}$ & $0.01506_{-0.00062}^{+0.0012}$\\
                     ~~~$T_{14}$\dotfill &Total duration (days)\dotfill & $0.212_{-0.022}^{+0.027}$    & $0.2156\pm0.0022$\\
\enddata                                                                                                 
\tablecomments{The best parameters for the star KELT-2B are in the text at the end of \S 3.}             
\end{deluxetable}  

\begin{figure}
\centerline{\includegraphics[scale=1.5]{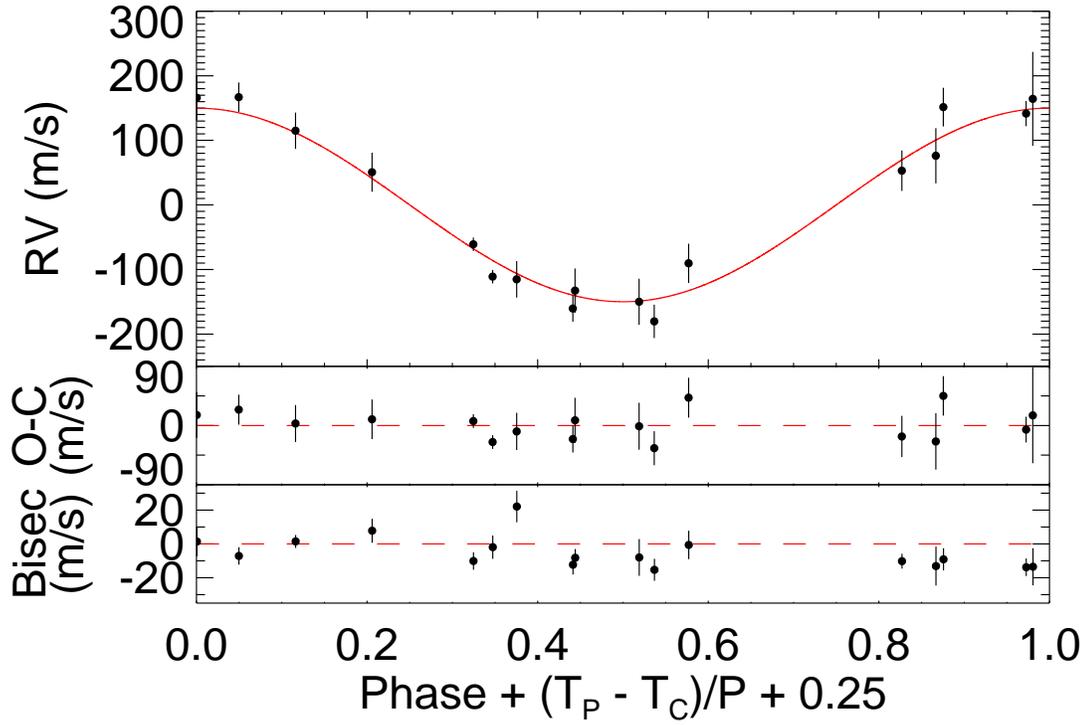}}
\caption{TRES radial velocity measurements of KELT-2A. The top panel shows the observations phased to our best orbital model, shown in red. We have removed the system's systemic velocity of 47.5 km s$^{-1}$ and the best fit slope. The middle panel shows the residuals of the radial velocity observations to our orbital fit, and the bottom panel shows the bisector span of each observations as a function of phase.}
\end{figure}

\clearpage

\begin{figure}
\centerline{\includegraphics[scale=1]{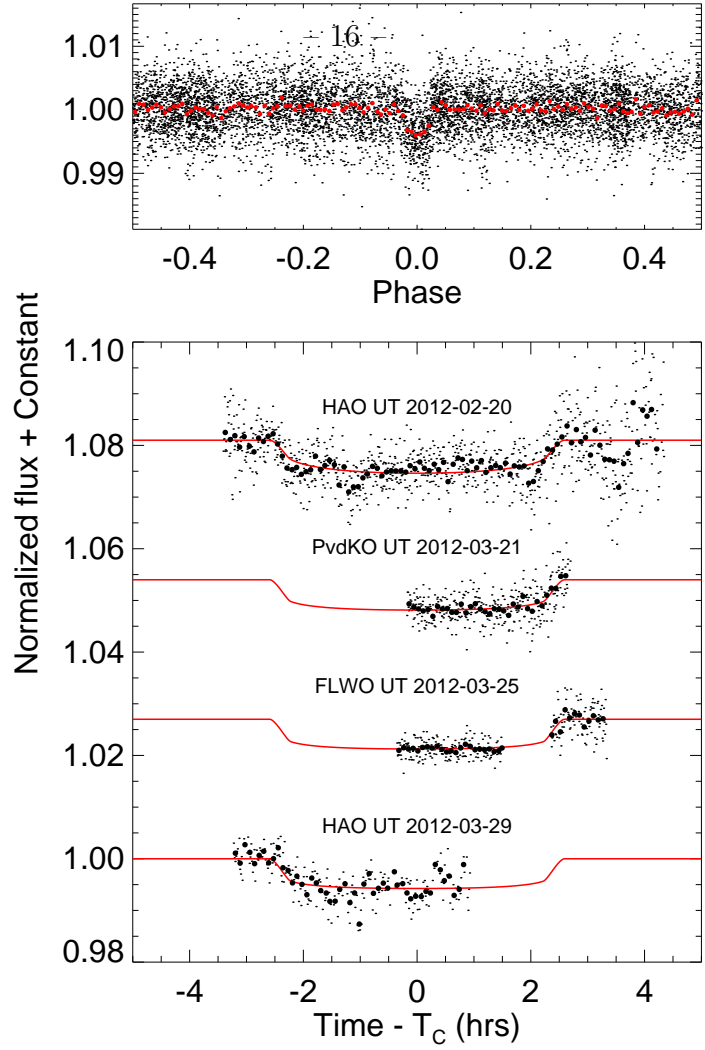}}
\caption{KELT-North discovery photometry (top panel), and then the follow-up transit observations (bottom panel) of KELT-2. The red overplotted line in the bottom panel is the best fit transit model. Note that all the follow-up lightcurves have been deblended, to remove the light from KELT-2B.}
\end{figure}

\newpage

\begin{figure}
\centerline{\includegraphics[scale=.75]{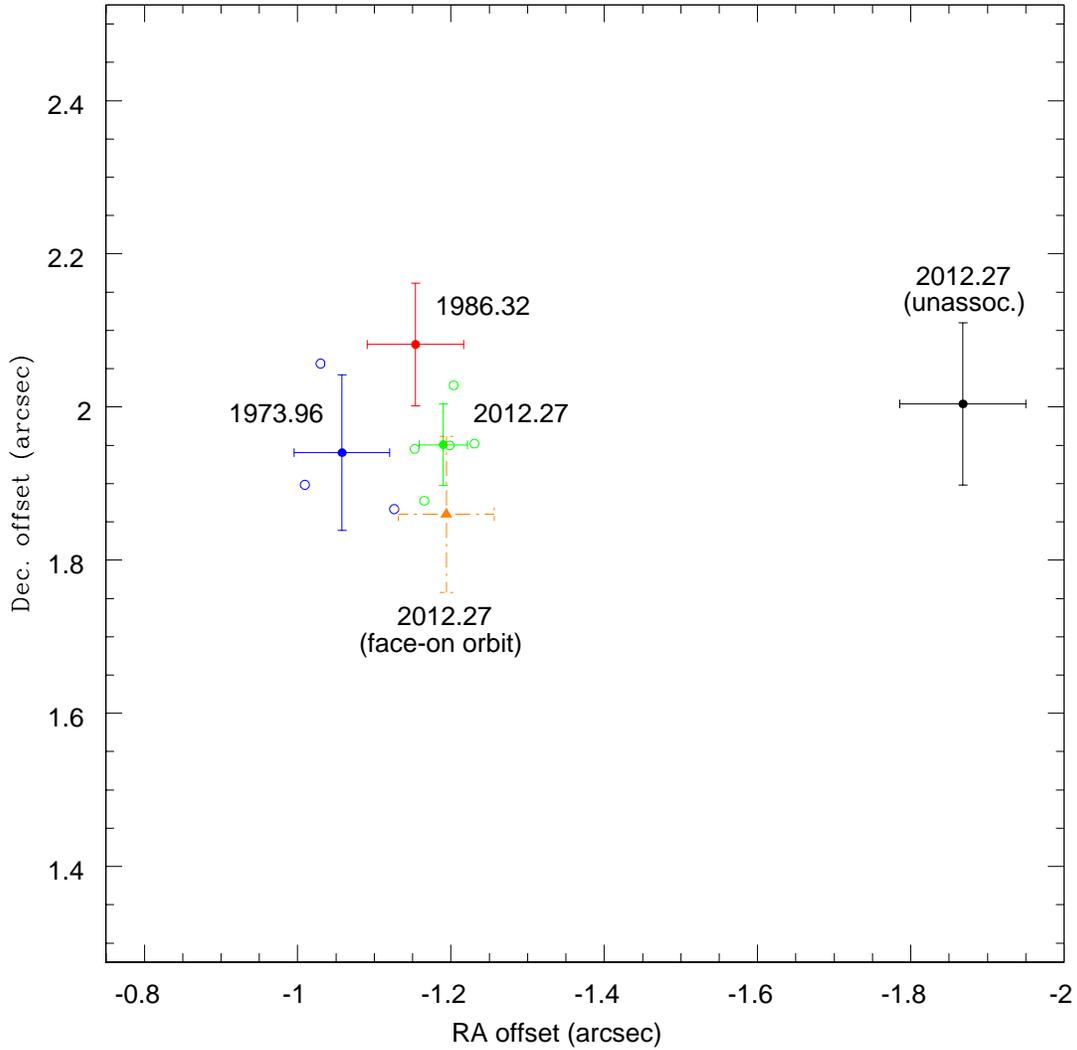}}
\caption{The position of KELT-2B relative to A. The open blue points are observations by \cite{couteau1975} -- the solid blue point with error bars is their average. The red point is from \cite{argue1992}. The open green points are the observations we made on UT 2012-04-10, and the solid green point with errors is their average. The black point shows the relative position of KELT-2B with respect to A if star B was an unassociated background star. The orange triangle shows where KELT-2B would be if it started at the 1973.96 position and were on a face-on circular orbit.}
\end{figure}

\begin{figure}
\centerline{\includegraphics[scale=0.75,angle=90]{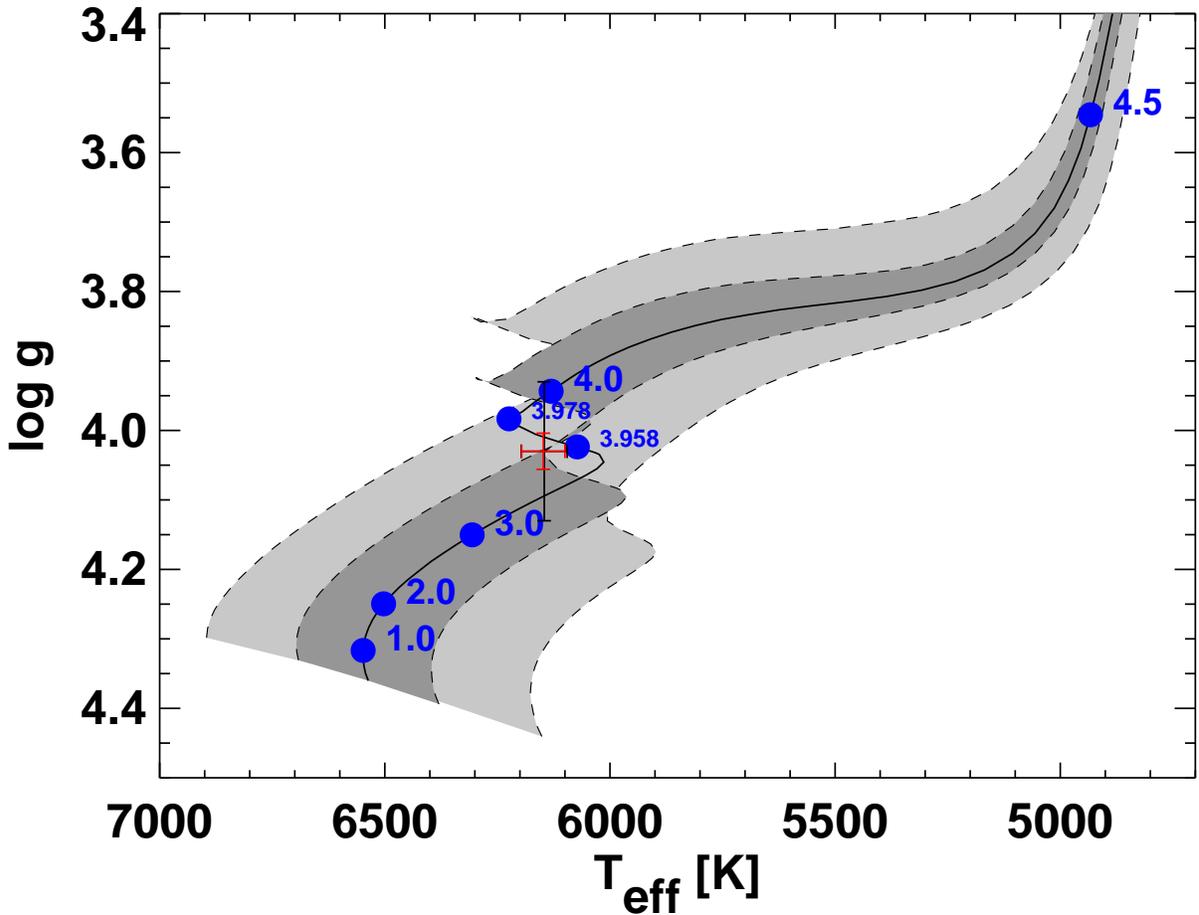}}
\caption{A theoretcial HR diagram showing the position of KELT-2A on the Yonsei-Yale models \citep{demarque2004}. The solid center track corresponds to our adopted parameters for KELT-2A described in \S3. Various ages, in Gyr, are marked in blue. The red cross shows our final constraints on $\teff$ and $\log(g)$ from the adopted {\sc exofast} results, while the black cross show the constraints from solely the spectroscopy. The dark and light shaded regions show the 1-$\sigma$ uncertainty arising from the uncertainties in our values of $M_*$ and [Fe/H] determined by {\sc exofast} (dark) and only spectroscopy (light).}
\end{figure}

\end{document}